\documentclass[prb,showpacs,preprintnumbers,amsmath,amssymb]{revtex4}
\usepackage{graphicx}%
\usepackage{color}

\begin{document}

\title{Relativistic effects and fully spin-polarized Fermi surface at the Tl/Si(111) surface}

\author{Julen Iba\~{n}ez Azpiroz$^{1,2}$, Asier Eiguren$^{1,2}$ and Aitor Bergara$^{1,2,3}$}
\address{$^{1}$Materia Kondentsatuaren Fisika Saila, Zientzia eta
Teknologia Fakultatea, Euskal Herriko Unibertsitatea, 644
Postakutxatila, 48080 Bilbao, Basque Country, Spain}
\address{$^{2}$Donostia International Physics Center (DIPC), Paseo Manuel de Lardizabal 4, 20018 Donostia/San Sebastian, Spain}
\address{$^3$Centro de F\'{i}sica de Materiales CFM - Materials Physics Center MPC, Centro Mixto CSIC-UPV/EHU,
Edificio Korta,
Avenida de Tolosa 72, 20018 Donostia, Basque Country, Spain}

\date{\today}


\begin{abstract}
We present a detailed analysis of the relativistic electronic structure and the momentum dependent spin-polarization of the Tl/Si(111) surface. Our first principle calculations reveal the existence of fully spin-polarized electron pockets associated to the huge spin-splitting of metallic surface bands. 
The calculated spin-polarization shows a very complex structure in the reciprocal space, 
strongly departing from simple theoretical model approximations.
Interestingly, the electronic spin-state close to the Fermi surface
is polarized along the surface perpendicular direction and reverses its orientation 
between different electron pockets.

\end{abstract}
\maketitle

\section{Introduction}
\label{intro}

The role played by the electronic spin in nominally non-magnetic surfaces has attracted a considerable interest in the last two decades due to the potential applications of these systems in the emergent field of spintronics~
\cite{datta,guimaraes_graphene-based_2010,nitta_gate_1997,awschalon}. 
Soon after the discovery of a spin-splitted surface band in the Au(111) metallic surface~\cite{lashell},
a large variety of surfaces has been intensively examined, both theoretically and experimentally~\cite{asier,bi100,sb111,dil_ag111,hw110}. It is now well established that the origin of the surface electron spin-splitting resides on the 
lack of the inversion symmetry close to the surface area. 
Interestingly, recent model calculations ~\cite{minghao,premper} suggest the important role played by the 
in-plane inversion asymmetry in the enhancement of the spin-splitting magnitude, as well as in the determination of the spin-polarization structure.

Among the many surfaces exhibiting spin-splitting phenomena,
semiconductor substrates covered by a single heavy-element overlayer
demonstrate specially encouraging properties for possible spintronic 
applications~\cite{meier_quantitative_2008,scontr}. 
On one hand, these type of surfaces have been recently found to exhibit exceptional 
relativistic effects~\cite{ast_giant_2007,bisi111}, inducing spin-orbit energy shifts two orders of magnitude bigger than those found at normal semiconductor heterojunctions. 
From the practical point of view, 
heavy elements such as Tl or Sb are widely used in electronic instruments like 
infrared detectors~\cite{Nayar:77} or Hall-effect devices~\cite{hallsb}. 
Another important property of semiconductor surfaces is the band gap 
associated to the semiconductor substrate, which ensures the
two-dimensional character of the transport properties and the absence of 
any appreciable bulk contribution,  
both conditions being indispensable for an effective manipulation of the surface spin-state.

The Tl/Si(111) surface is an outstanding example in which 
the spin-orbit interaction plays a critical role in determining the transport properties of the system. 
The honeycomb layered structure of the Si(111) substrate induces a singular spin-pattern in momentum space that departs from simple pictures such as the Rashba model~\cite{rashba}. This peculiar property is farther enhanced by the strong relativistic effects associated to the Tl overlayer, producing a highly complex spin-configuration in the neighborhood of the Fermi level. In this article, we investigate the nature of the spin-orbit interaction on two different terminations of the Tl/Si(111) surface describing the relativistic electronic structure and the momentum dependent spin-polarization. 

\begin{figure}
   \centering
\includegraphics[width=1.0\textwidth]{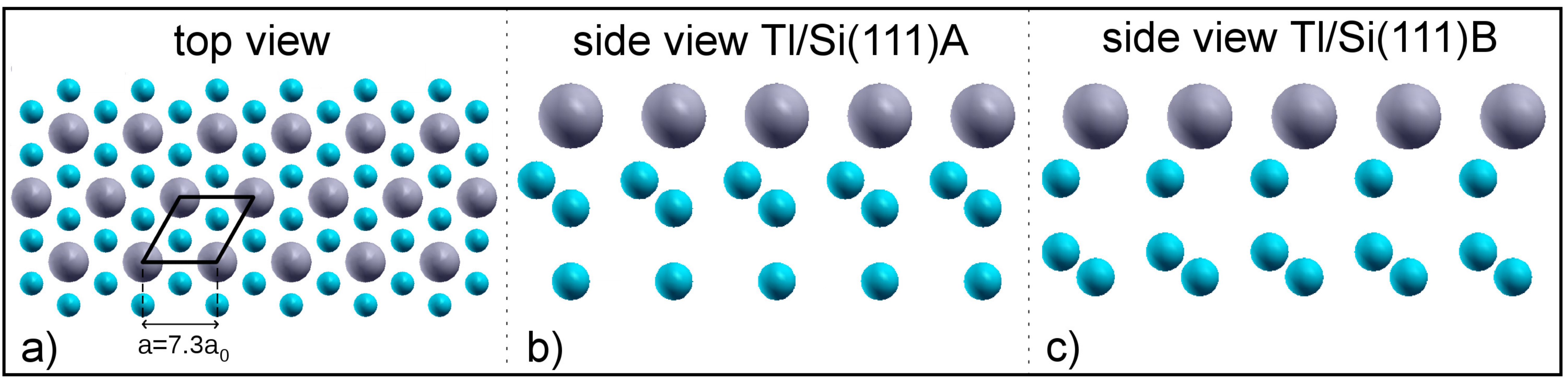}
\caption{(color online) (a) Top view of the Tl/Si(111) surface. Big (gray) spheres represent the Tl surface monolayer, while the small (blue) ones are the Si substrate layers. The solid (black) lines denote the projection of the surface unit cell. (b) and (c) Side view of the two surface terminations considered, Tl/Si(111)A and Tl/Si(111)B.}
\label{structures}
\end{figure}

\section{Computational method}
\label{cd}

We have considered the non-collinear DFT formalism and plane waves as basis functions for the expansion of the Kohn-Sham orbitals~\cite{espresso,Dalcorso}. The convergence of the plane wave basis has been achieved with an energy cutoff of 50 Ry. The integrations over the surface Brillouin zone have been performed using the tetrahedron method~\cite{tetrahedra} considering a $32\times32$ Monkhorst-Pack mesh~\cite{MParck}. The exchange-correlation energy has been approximated within the PBE-GGA parametrization~\cite{pbe,scuseria}.

We used norm-conserving fully relativistic pseudopotentials as illustrated in refs.~\cite{bylander,ppteurich}. These pseudopotentials describe the relativistic effects up to order $1/c^{2}$, including the mass-velocity, the Darwin and the spin-orbit coupling terms~\cite{kleinman}. The electron wave functions are treated within the spinor formalism in order 
to properly incorporate the non-collinear effects associated to the spin-orbit interaction.

In this article, we consider two different surface terminations designated as Tl/Si(111)A and Tl/Si(111)B (see Fig. \ref{structures}). Both terminations were modeled within the repeated slab approach consisting of 40 atom layers, imposing a relaxation criterion of $|F_{i}|<$10$^{-4}$ Ry au$^{-1}$. 
Previous experimental and theoretical investigations ~\cite{t42,adsorption-ener} demonstrated that the Tl/Si(111)A termination is the most stable. However, in our 
calculations we find that the Tl/Si(111)B termination is energetically competitive,
with a tiny energy cost of $\Delta E_s$ $\simeq10^{-3}$ Ry per Tl atom.

\section{Results}
\label{results}

\subsection{Tl/Si(111)A}
\label{term_a}

\begin{figure}
   \centering
\includegraphics[width=0.7\textwidth]{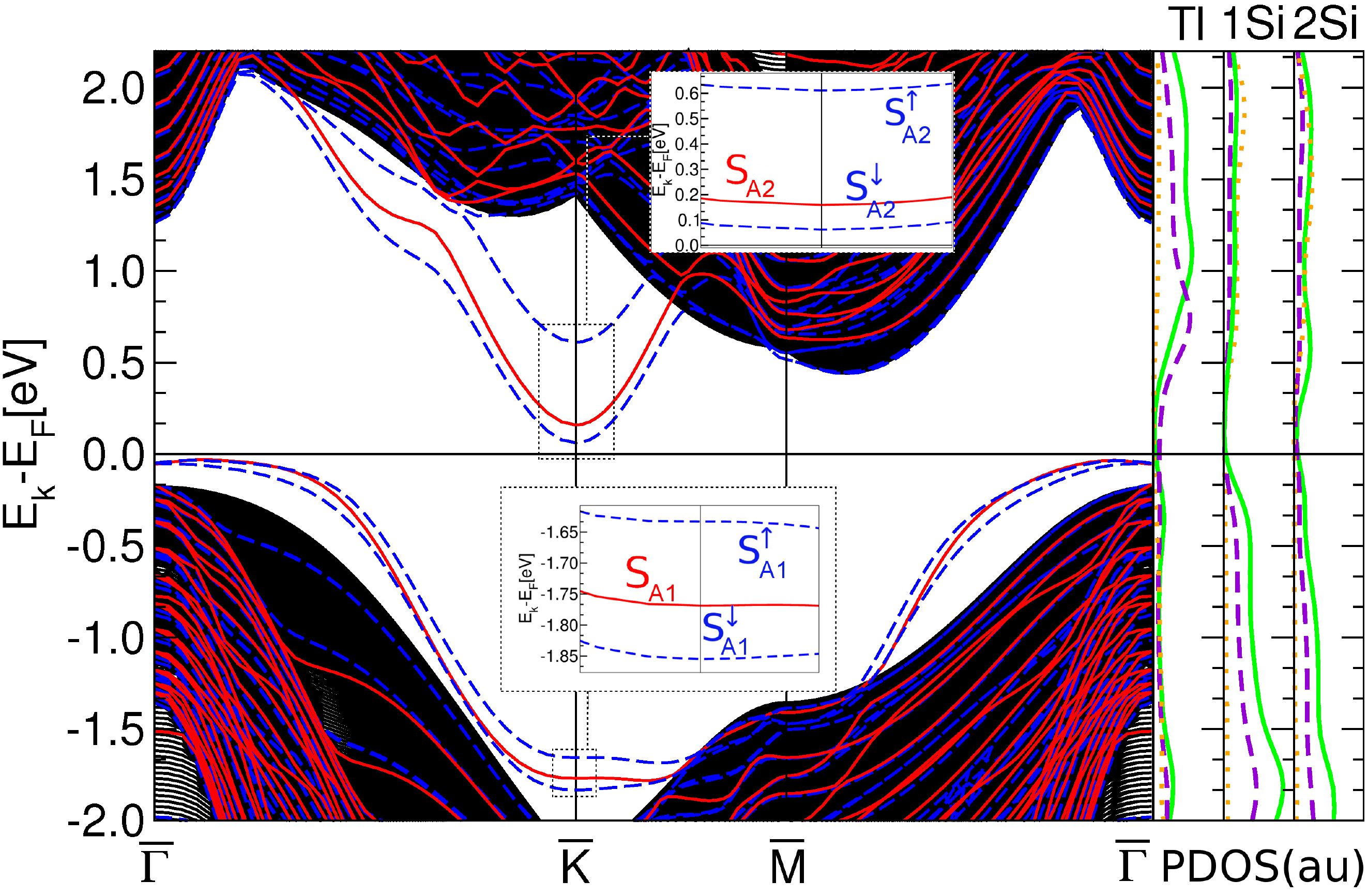}
\caption{(color online) (Left) Band structure of the Tl/Si(111)A surface termination. 
The scalar relativistic and fully relativistic bands are represented by solid (red) 
and dashed (blue) lines, respectively. 
The continuous background denotes the bulk band projection. 
Inset figures show the details of the surface bands in the neighborhood of 
high symmetry point $\overline{K}$. 
(Right) Projected DOS for the Tl surface monolayer and the first two Si layers. 
$np_{3/2}$, $np_{1/2}$ and $ns_{1/2}$ orbitals 
(principal quantum number $n=6$ for Tl, $n=3$ for Si) 
are represented by solid (green), dashed (violet) 
and dotted (orange) lines, respectively.}
\label{A-band}
\end{figure}

Fig. \ref{A-band} illustrates our results for the electronic structure of the Tl/Si(111)A termination. 
The scalar relativistic calculation (without spin-orbit term) produces two surface states, 
labeled as $S_{A1}$ and $S_{A2}$ (see Fig. \ref{A-band}). 
This calculation predicts a semiconductor state with an energy gap of approximately 0.2 eV, since
neither $S_{A1}$ nor $S_{A2}$ bands cross the Fermi level.
In contrast, the fully relativistic calculation presents four surface bands labeled
as $S^{\downarrow}_{A1}$, $S^{\uparrow}_{A1}$, $S^{\downarrow}_{A2}$ and $S^{\uparrow}_{A2}$. 
These bands are interpreted as originating from the spin-splitting of the scalar relativistic $S_{A1}$ and $S_{A2}$ bands.

It is evident from Fig. \ref{A-band} that the spin-orbit interaction induces 
a considerable perturbation on the surface bands associated to the scalar relativistic calculation.
The spin-degeneracy of the surface bands 
at the $\overline{\Gamma}$ and $\overline{M}$ points (Fig. \ref{A-band}) is a consequence of the
combination of the C3 rotational and the time reversal symmetry of the system~\cite{minghao}.
In contrast, these symmetry considerations do not forbid a finite spin-orbit energy shift at high symmetry point $\overline{K}$. The inset figures of Fig. \ref{A-band} reveal the exact magnitude of the spin-orbit interaction close to $\overline{K}$ point, finding that the $S^{\downarrow}_{A1}$ and $S^{\uparrow}_{A1}$ bands are spin-split by approximately 0.25 eV, in good agreement 
with ARPES photoemission measurements~\cite{abrupt}. 
In an analogous way, the $S^{\downarrow}_{A2}$ and $S^{\uparrow}_{A2}$ bands suffer
a maximum splitting of $\sim$ 0.6 eV, an
extraordinarily large value for an spin-orbit energy shift. 
The energy band gap of this termination reduces roughly 
from a value of $0.2$ eV in scalar relativistic calculations to the $0.1$ eV found in fully relativistic bands.

\begin{figure}

   \centering
\includegraphics[width=0.7\textwidth]{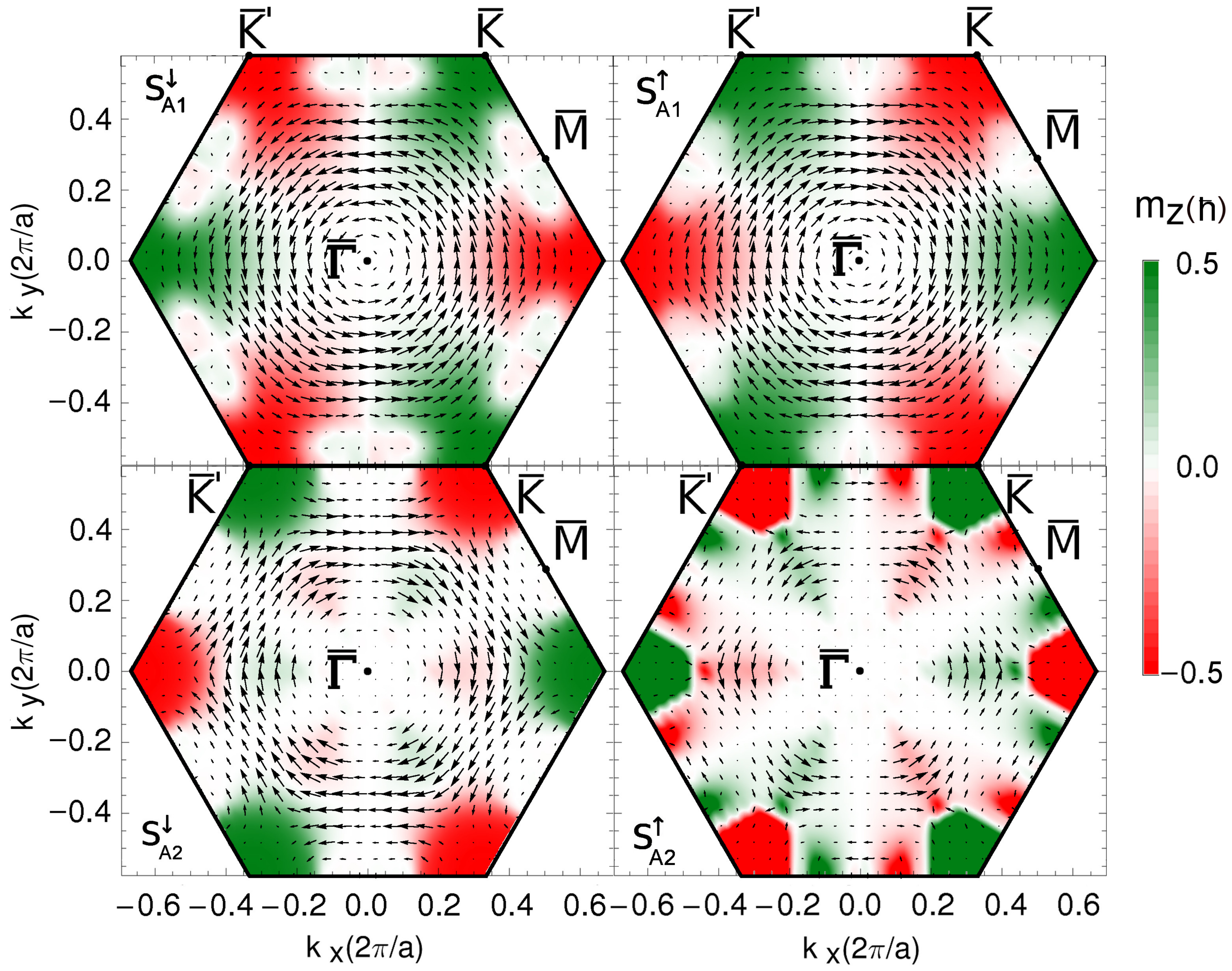}
\caption{(color online) Spin-polarized structure through the entire surface 
Brillouin zone. Arrows (black) represent the in-plane spin-polarization component, 
while the background reflects the surface perpendicular 
component $m_{z,i}(\textbf{k})$ (the scale ranges $[-0.5\hbar,0.5\hbar]$).}
\label{A-spin-pol}
\end{figure}
\begin{figure}
   \centering
\includegraphics[width=0.6\textwidth]{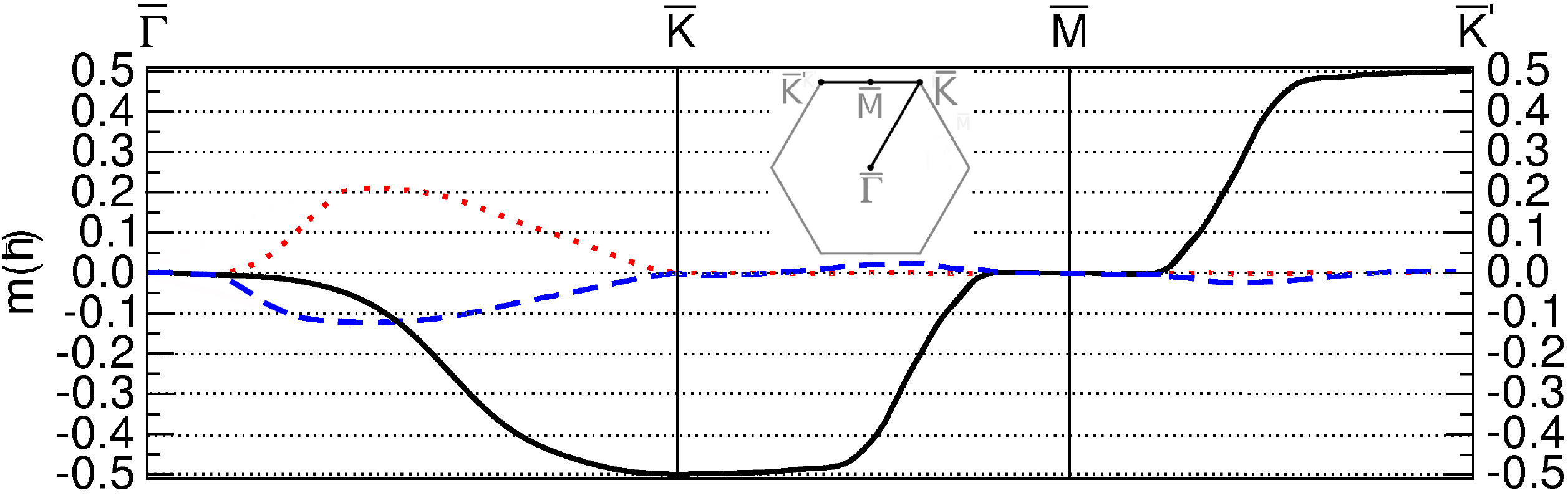}
\caption{(color online) The calculated spin-polarization components 
$m_{x,S^{\downarrow}_{A2}}(\textbf{k})$, $m_{y,S^{\downarrow}_{A2}}(\textbf{k})$ 
and $m_{z,S^{\downarrow}_{A2}}(\textbf{k})$ are represented along the high 
symmetry line $\overline{\Gamma}-\overline{K}-\overline{M}-\overline{K^{\prime}}$ 
by dotted (red), dashed (blue) and solid (black) lines, respectively.} 
\label{A-spin-pol-path}
\end{figure}

In Fig. \ref{A-spin-pol} we present the calculated momentum dependent spin-polarization for the four spin-splitted states. The spin-polarization is defined as the expectation value of the Pauli matrices  
\begin{eqnarray}\label{spin-pol}
m_{\alpha,i}({\bf k})=\dfrac{1}{\Omega}\sum_{\sigma\sigma^{\prime}}\int
 \phi^{\sigma^{\prime}*}_{{\bf k},i}({\bf r})  \sigma_{\alpha}^{\sigma^{\prime}\sigma}
\phi^{\sigma}_{{\bf k},i}({\bf r})\mbox{d}\bf{r},
\end{eqnarray}
where $\Omega$ denotes the volume of the system and $\alpha$ runs over the cartesian axes.
$\phi^{\sigma}_{{\bf k},i}({\bf r})$ represent the Kohn-Sham eigen-spinor of the surface 
states, while $\sigma_{\alpha}^{\sigma^{\prime}\sigma}$ denote the matrix elements of the Pauli spin-operator. 
As demonstrated in Fig. \ref{A-spin-pol}, a given surface state is spin-polarized in
approximately the opposite direction with respect to its associated spin-splitted state. 
The negligible spin-polarization around high symmetry 
points $\overline{\Gamma}$ and $\overline{M}$ is consistent 
with the null spin-splitting observed in the 
band structure in these regions. 

It is commonly accepted, on the grounds of the Rashba model~\cite{rashba}, 
that the spin-state of the surface electrons 
is constrained to lie parallel to the surface plane. 
Fig. \ref{A-spin-pol} depicts an in-plane rotational spin-polarization 
around the $\overline{\Gamma}$ point, qualitatively resembling the Rashba picture. In addition,
Fig. \ref{A-spin-pol} reveals that close to the $\overline{K}$ 
and $\overline{K^{\prime}}$ symmetry points, 
the spin-state of the surface electrons becomes predominantly polarized along the $z$ direction,
in agreement with recent spin-resolved ARPES measurements~\cite{abrupt}.
This characteristic property is a consequence of the C3 rotational symmetry of the honeycomb layered structure of the surface~\cite{minghao}.

In Fig. \ref{A-spin-pol-path} we present a quantitative analysis of the spin-polarization components of the $S^{\downarrow}_{A2}$ band along the $\overline{\Gamma}-\overline{K}-\overline{M}-\overline{K^{\prime}}$ high symmetry lines.
These results demonstrate that in the neighborhood of 
high symmetry points $\overline{K}$ and $\overline{K^{\prime}}$
the absolute value of the $z$ component almost reaches the maximum value, 
0.5$\hbar$, while the in-plane components become negligible. 
Furthermore, Fig. \ref{A-spin-pol-path} indicates that the electronic states 
around high symmetry points $\overline{K}$ and $\overline{K^{\prime}}$ are 
spin-polarized in completely opposite directions.
These remarkable properties make the Tl/Si(111)A termination a particularly interesting system to study for instance, the low energy transport properties or
optically induced spin-flip transitions~\cite{spin-flip}.

\begin{figure}
   \centering
\includegraphics[width=0.7\textwidth]{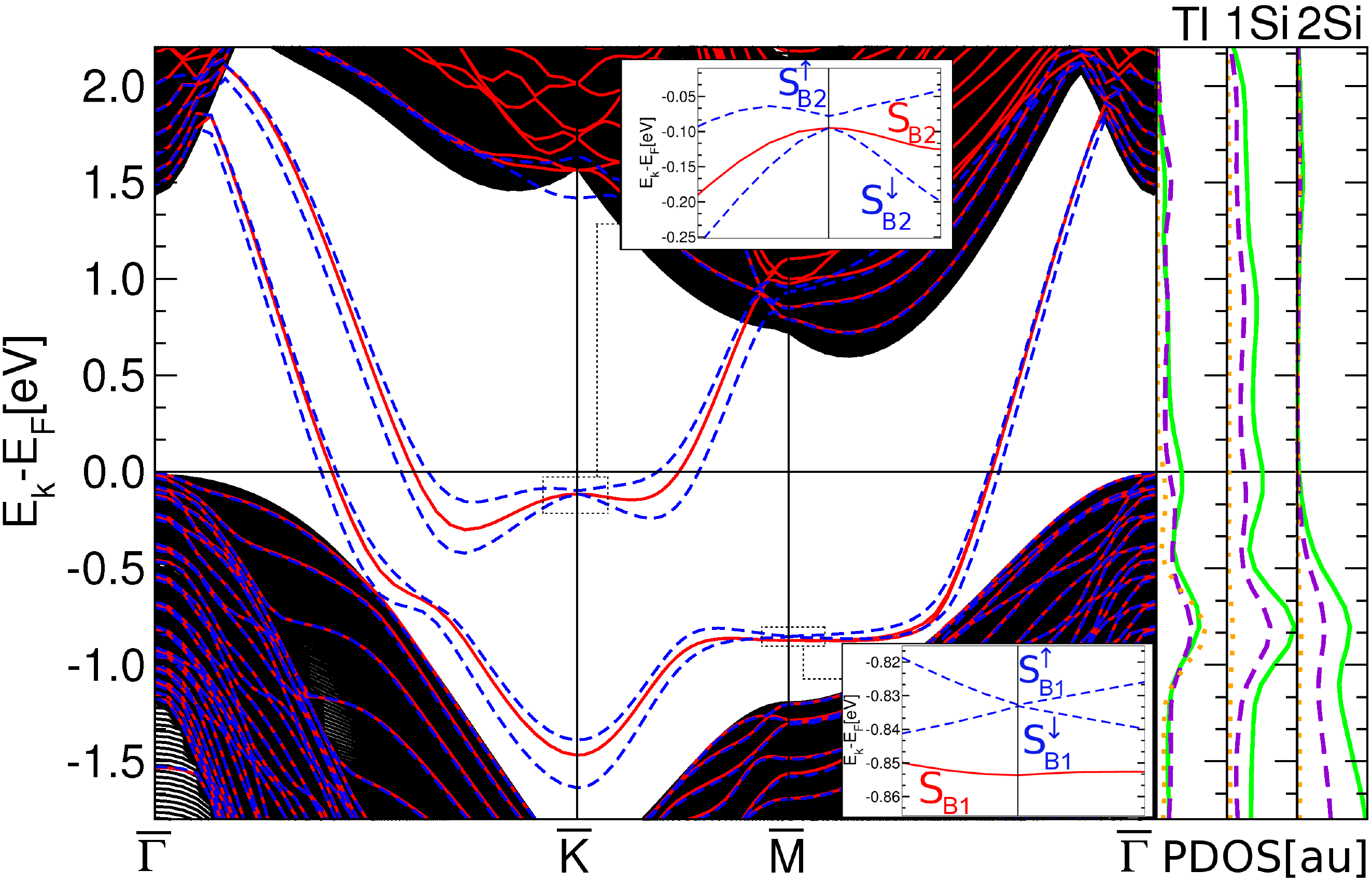}
\caption{(color online) (Left) Band structure of the Tl/Si(111)B surface termination. The scalar relativistic and fully relativistic bands are represented by solid (red) and dashed (blue) lines, respectively. The continuous background denotes the bulk band projection. Inset figure connected to the $\overline{M}$ point shows a complete spin-degeneracy of $S^{\downarrow}_{B1}$ and $S^{\uparrow}_{B1}$ bands. Inset figure connected to $\overline{K}$ point reveals a quasi-degenerate configuration of $S^{\downarrow}_{B2}$ and $S^{\uparrow}_{B2}$ bands ($\Delta E_{s}\sim25$ meV). The Fermi surface of the system is included as an inset figure, middle left. (Right) Projected DOS for the Tl surface monolayer and the first two Si layers. $np_{3/2}$, $np_{1/2}$ and $ns_{1/2}$ orbitals (principal quantum number $n=6$ for Tl, $n=3$ for Si) are represented by solid (green), dashed (violet) and dotted (orange) lines, respectively. The energy regions around $-$0.75 and $-$0.1 eV show non-negligible Tl $6s$ orbital contribution.}
\label{B-band}
\end{figure}

\subsection{Tl/Si(111)B}
\label{term_b}

\begin{figure}
   \centering
\includegraphics[width=0.35\textwidth]{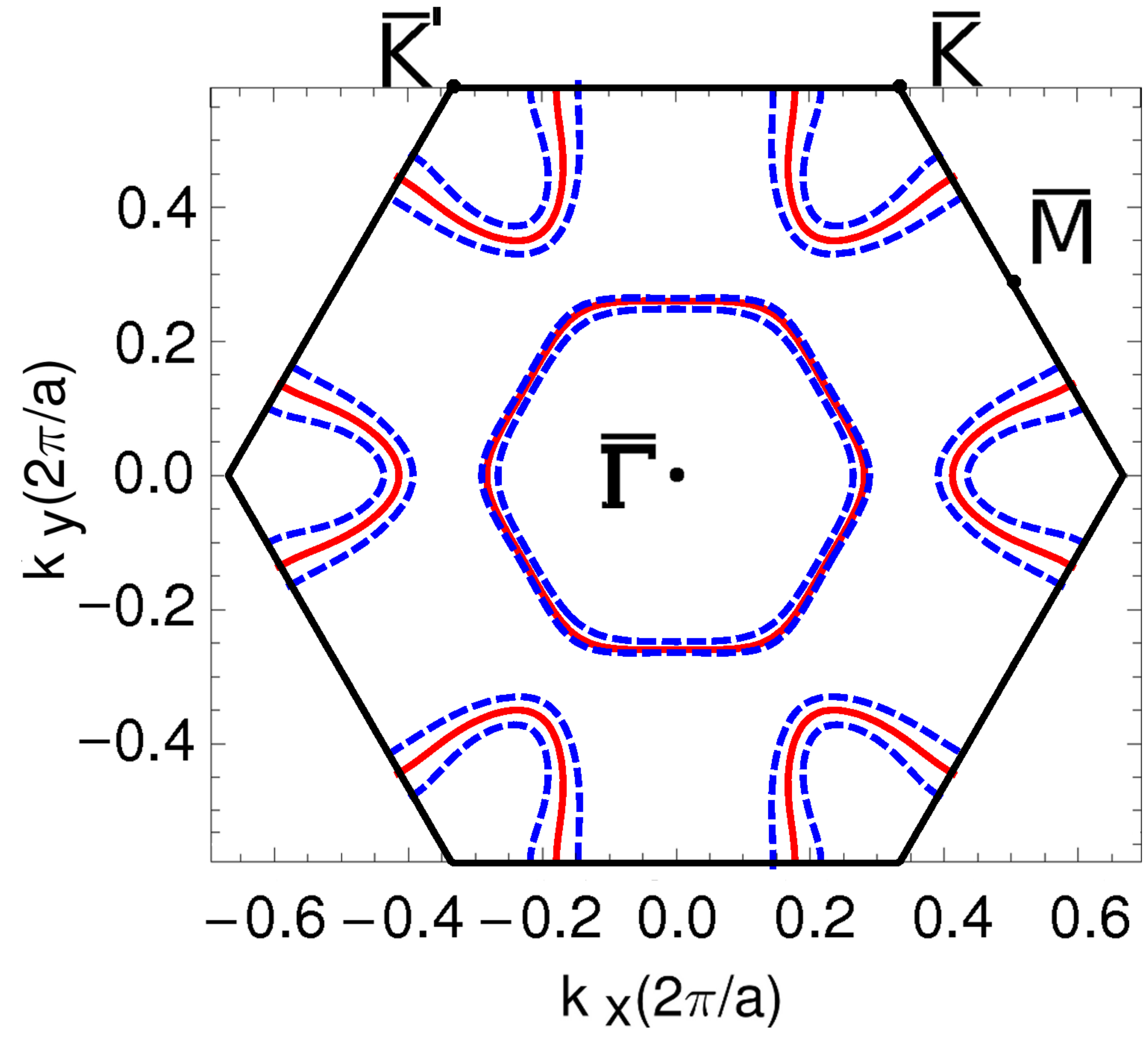}
\caption{(color online) Fully spin-polarized Fermi surface of the Tl/Si(111)B 
surface termination. Solid (red) and dashed (blue) lines represent the Fermi crossing points 
of scalar relativistic and fully relativistic surface bands, respectively.
The inner and outer electron pockets around $\overline{K}$ and $\overline{K^{\prime}}$ points 
belong
to $S^{\uparrow}_{B2}$ and $S^{\downarrow}_{B2}$ bands, while  
the inner and outer electron-hole pockets around the $\overline{\Gamma}$ point belong
to $S^{\downarrow}_{B1}$ and $S^{\uparrow}_{B1}$ bands, respectively.}
\label{FS}
\end{figure}

The electronic structure of the Tl/Si(111)B termination is presented in Fig. \ref{B-band}. 
In this termination, we find that four spin-splitted surface bands 
($S^{\downarrow}_{B1}$, $S^{\uparrow}_{B1}$, $S^{\downarrow}_{B2}$ and $S^{\uparrow}_{B2}$) 
cross the Fermi level, producing a fully spin-polarized Fermi surface (Fig. \ref{FS}).
The $S^{\downarrow}_{B2}$ and $S^{\uparrow}_{B2}$ bands form several 
spin-polarized electron pockets around the high symmetry points $\overline{K}$ and $\overline{K^{\prime}}$.
The $S^{\downarrow}_{B1}$ and $S^{\uparrow}_{B1}$ bands are occupied all over the Brillouin zone except around 
high symmetry point $\overline{\Gamma}$, where we find an electron-hole pocket of radius $k_{F}\sim 0.46$  $\text{\AA}^{-1}$ 
(see Figs \ref{B-band} and \ref{FS}). 
Consequently, the Tl/Si(111)B termination exhibits a strong metallic character 
entirely induced by the fully relativistic surface bands.

The $S^{\downarrow}_{B1}$ and $S^{\uparrow}_{B1}$ states
are maximally spin-split close to the $\overline{K}$ point ($\sim$ 0.25 eV).
These bands become spin-degenerate at the $\overline{M}$ point, as it can be appreciated in the inset of Fig. \ref{B-band}.
Similarly, the overall spin-splitting for the $S^{\downarrow}_{B2}$ and $S^{\uparrow}_{B2}$ surface bands is found to be of the order of 0.2 eV. Close to the $\overline{K}$ point, these bands
become accidentally quasi-degenerate and the magnitude of 
the splitting diminishes to a negligible but finite value of $\sim$ 25 meV (inset of Fig. \ref{B-band}).

\begin{figure}[t]
   \centering
\includegraphics[width=0.7\textwidth]{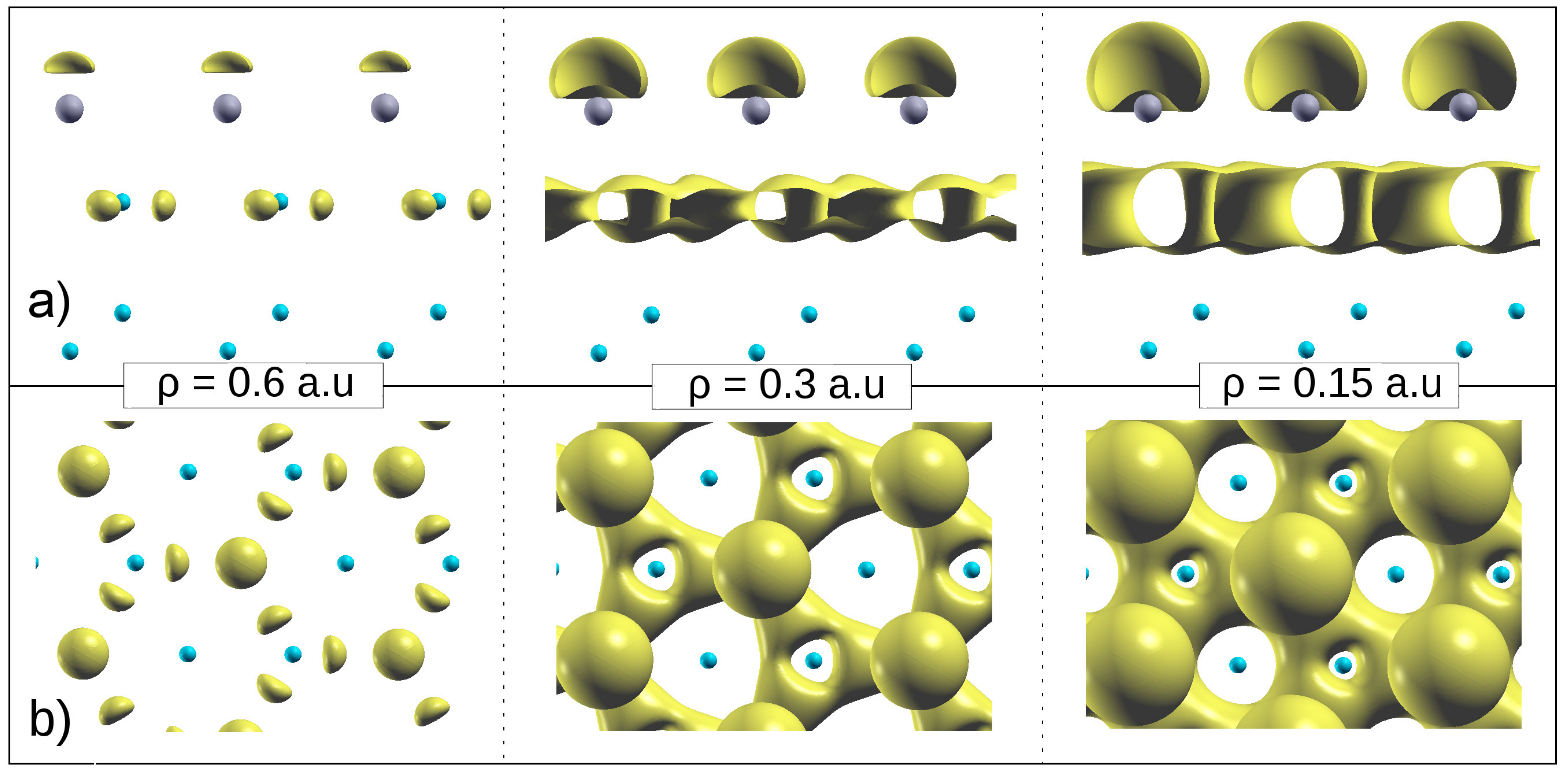}
\caption{(color online) Charge distribution of $S^{\downarrow}_{B2}$ state at high symmetry point $\overline{K}$. Big (gray) and small (blue) spheres represent Tl and Si atoms, respectively. (a) and (b) illustrate the side and top views of the charge density isosurfaces corresponding to $\rho=$ 0.6, 0.3 and 0.15 a.u. The shape of the charge distribution around Tl atoms shows a 'front lobe' associated to a $sp_{z}$ hybrid orbital with a predominant $s$ to $p_{z}$ ratio. An almost identical picture is obtained for the $S^{\uparrow}_{B2}$ state.}
\label{B-wf}
\end{figure}
The $s$ orbital character of the surface electronic wave functions is indicative of a possible spin-degeneracy. The right panel of Fig. \ref{B-band} shows the projected density of states~\cite{daniel} (PDOS) for various orbital components.
We find that the Tl $6s$ orbitals represent the largest contribution to the PDOS at approximately $-$0.75 eV. 
Similarly, we find a non-negligible contribution of these orbitals at around $-$0.1 eV. 
These two energy regions with
non-negligible Tl $6s$ contribution coincide with the energy regions of the inset figures of Fig. \ref{B-band}.

Fig. \ref{B-wf} illustrates several charge 
isosurfaces ($\rho = \sum_{\sigma} | \phi^{\sigma}_{\textbf{k},i}({\bf r})|^{2}$) 
associated to the $i=S^{\downarrow}_{B2}$ state at high symmetry point $\overline{K}$. 
As demonstrated in the figure, this surface state is localized within the first two layers of the slab. 
Close to the Tl atoms, where relativistic effects prevail, the charge distribution
shows a characteristic 'front lobe' shape associated to an atomic $sp_{z}$ hybrid orbital.
The big isosurface volume of the 'front lobe' indicates that 
the $s$ character predominates over the $p_{z}$. 
Interestingly, close to the $\overline{M}$ point, very similar charge distributions are 
found for the $S^{\downarrow}_{B1}$ and $S^{\uparrow}_{B1}$ states, concluding that, effectively,
the Tl $6s$ character is predominant in the spin-degenerate regions.

\begin{figure}[t]
   \centering
\includegraphics[width=0.7\textwidth]{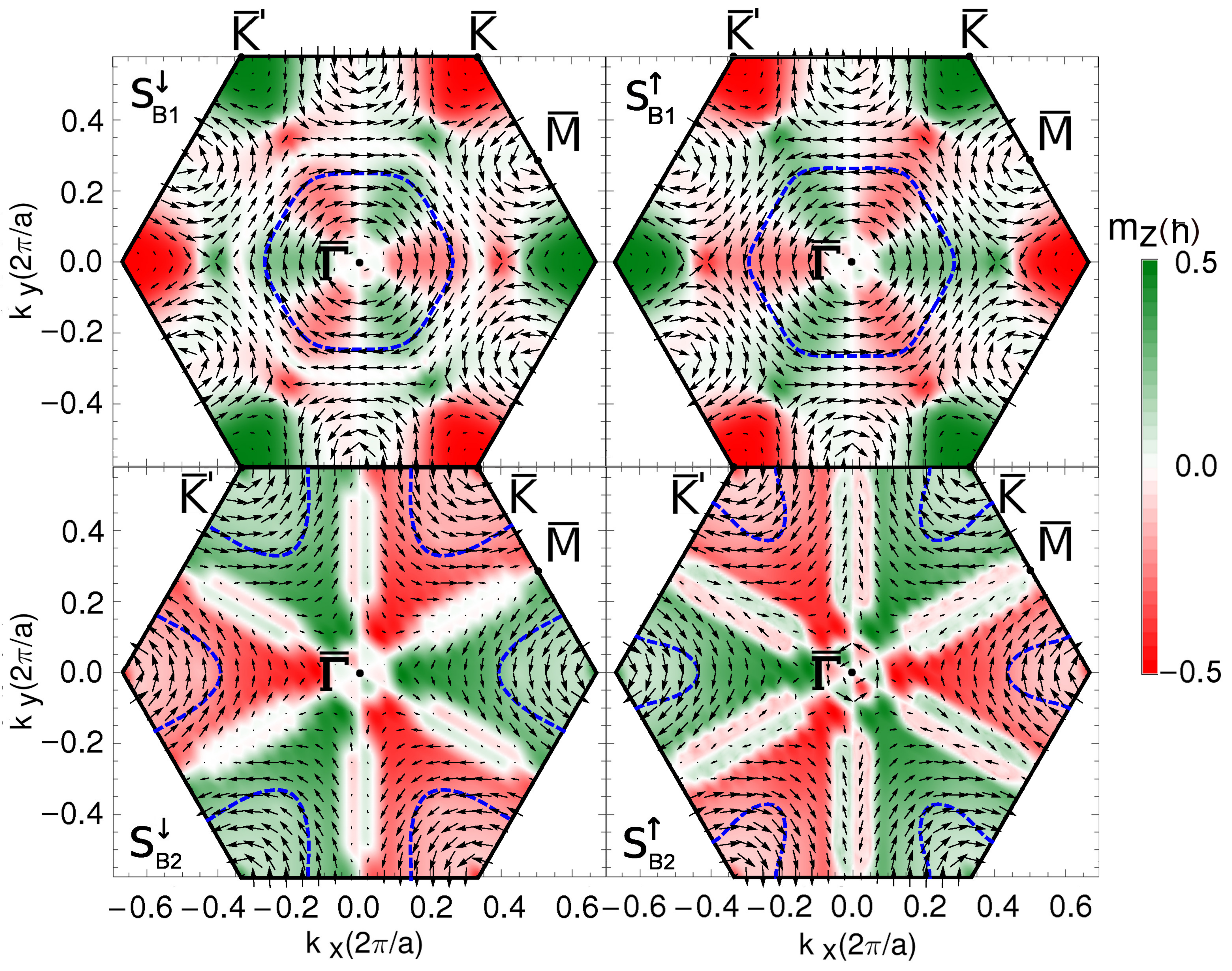}
\caption{(color online) Spin-polarized structure through the entire surface 
Brillouin zone. Arrows (black) represent the in-plane spin-polarization component, 
while the background reflects the surface perpendicular component 
$m_{z,i}(\textbf{k})$. 
The Fermi surface for each state is depicted by the dashed (blue) lines. 
}
\label{B-spin-pol}
\end{figure}

Fig. \ref{B-spin-pol} presents the calculated spin-polarization for the different 
surface states of the Tl/Si(111)B termination, exhibiting 
a far more complex structure than in the previous termination. 
In agreement with symmetry considerations, the spin is found to be
100$\%$ polarized along the surface perpendicular direction at high symmetry points $\overline{K}$ and $\overline{K^{\prime}}$, its orientation
being reversed going from one point to the other.

Our ab-initio calculations demonstrate an important contribution of the
$z$ spin-polarized component over the entire Brillouin zone, 
specially for the $S^{\downarrow}_{B2}$ and $S^{\uparrow}_{B2}$ states. 
This behavior strongly departs from simple models calculations predicting 
a surface perpendicular spin-polarization only in a very small area around
$\overline{K}$ and $\overline{K^{\prime}}$~\cite{minghao}.
The calculated Fermi surface extends over the regions where the in-plane spin-polarization
is combined with an important surface perpendicular contribution (see Fig. \ref{B-spin-pol}). 
Interestingly, the spin-polarization reverses its orientation over the different electron pockets, 
providing the Tl/Si(111)B termination with unique transport properties.
A qualitative aspect revealed by these calculations is that the spin
is rotational and encircling an appreciable area around all the high symmetry points of the 
Brillouin zone ($\overline{\Gamma}$, $\overline{M}$ and $\overline{K}$). 
The spin-polarization structure in Fig. \ref{B-spin-pol} 
evidences that the details of a surface termination are sufficient to produce 
highly complex spin-patterns in reciprocal space, beyond simple theoretical models. 

\section{Conclusions}

In summary, we analyze the relativistic electron and spin-structure 
of two different terminations of the Tl/Si(111) surface, Tl/Si(111)A and Tl/Si(111)B. 
The calculations on the A termination are 
in very good agreement with spin-resolved ARPES photoemission experiments~\cite{abrupt}.
Our analysis demonstrates that the band gap of this surface is reduced to a value of $\sim$0.1 eV as a direct consequence of the spin-orbit interaction, the Fermi level being completely 
surrounded by spin-polarized states.
In overall, both terminations show a highly complex spin-polarization structure in momentum space,
particularly the Tl/Si(111)B termination. 
For this surface, we find that several spin-rotational centers are present, 
and that the in-plane spin-polarization
is combined with a substantial surface perpendicular component over the entire Brillouin zone.
All these features strongly depart from simple model theoretical predictions. 
The Tl/Si(111)B termination possess a strong metallic
character entirely induced by the relativistic surface bands and 
the calculated Fermi surface is mainly constituted by fully spin-polarized electron pockets around
high symmetry points $\overline{K}$ and $\overline{K^{\prime}}$. It is found that the spin-polarization reverses its orientation between different electron pockets and thus remarkable transport properties should be expected for both surfaces.

\section*{Acknowledgments}

The authors are grateful to M. Martinez-Canales, B. Rousseau and I. Errea for 
fruitful discussions, and acknowledge financial support from UPV/EHU 
(Grant No. IT-366-07) and the Spanish Ministry of Science and Innovation 
(Grant No. FIS2010-19609-C02-00). Computer facilities 
were provided by the Donostia International Physics Center (DIPC).

\section*{References}

%

\end{document}